# Sutte Indicator: an approach to predict the direction of stock market movements

Ansari Saleh Ahmar

*Department of Statistics, Faculty of Mathematics and Natural Sciences,
Universitas Negeri Makassar, Makassar City, 90223 Indonesia*

**Abstract**

The purpose of this research is to apply technical analysis of Sutte Indicator in stock trading which will assist in the investment decision making process i.e. buying or selling shares. This research takes data of "A" on the Indonesia Stock Exchange (IDX or BEI) 29 November 2006 until 20 September 2016 period. To see the performance of Sutte Indicator, other technical analysis are used as a comparison, Simple Moving Average (SMA) and Moving Average Convergence/Divergence (MACD). To see a comparison of the level of reliability prediction, the stock data were compared using *the mean absolute deviation* (MAD), *mean of square error* (MSE), and mean absolute percentage error (MAPE). The result of this research is that Sutte Indicator can be used as a reference in predicting stock movements, and if it is compared to other indicator methods (SMA and MACD) via MAD, MSE, and MAPE, the Sutte Indicator has a better level of reliability.

**Keywords:** stock market, Sutte Indicator, technical analysis, predicting

## 1. Introduction

The current economic growth in a country is very important. Economic growth is important because with this economic growth it can be said that the country has a good economic situation. In addition, good economic growth will be an attraction for investors to invest in the country. Investing an investor of various forms, such as investment in infrastructure development or it could be in the form of planting shares in a company. In stock trading, an investor is required to have minimal capability in stock trading, one of which is to know the movement of stock and market conditions. Both of these capabilities are very important for an investor in order to obtain maximum profit and minimum losses. If an investor does not have this capability then the investor will suffer losses due to his ignorance regarding the movement of stocks and market conditions. In stock trading, an investor usually uses a technical indicator in predicting stock movements. Technical indicators are usually graphic images showing the condition of stock movement. The main components that are often used in determining the movement of stocks are the opening price, closing price, the highest price, the lowest price, and the transaction volume. These components are often used in stock technical analysis. In determining stock movement, technical analysis has 2 categories namely lagging indicator and leading indicator. Lagging indicator is often used to measure a stock's trend and leading indicator is used to predict when it is overbought and oversold.

The movement of stocks in the stock market is not always consistent. It sometimes suddenly decreases and sometimes suddenly rises. One way to know the movement of a stock is by doing the technical analysis. Technical analysis or commonly called technical indicator has been developed by researchers in the field of economics such as Moving Average (Han, Yang, & Zhou, 2013), Relative Strength Index (RSI) (Abbey & Doukas, 2012), Stochastic, MAD, and Bolinger Band (Nithya & Thamizhchelvan, 2014). However, some of these technical indicators are less precise in predicting stock movements. Therefore, a new technical indicator is developed that is Sutte Indicator with the hope that this indicator can fix the problem.

*Corresponding author
  Email address: ansarisaleh@unm.ac.id

Sutte Indicator is developed by using four basic components of technical analysis i.e. opening price, closing price, highest price, and lowest price (Ahmar, Ahmar, Arifin, & Rahman, 2017) and also Sutte Indicator is a modification of the Moving Average indicator. In predicting stocks, Sutte Indicator uses two graphs that will show when it is the right time to make a purchase or sale of shares. From this chart, it will also give a signal to investors in order to obtain maximum profit with minimum losses.

## 2. Materials and Methods

In today's developmental era, many traders are experts in technical analysis and have the ability to modify technical indicators. Technical analysis is intended to forecast the price of a stock. The buyer and the seller set the price of the agreement that is fit, weighted and express all factors, quantitative and non-quantitative, rational and irrational, and is the only desirable picture (Suresh, 2013). Furthermore, Neely, and Weller (2011) said that technical analysis is the use of past price movements and / or other market data, such as volume of transaction, to assist in the decision making process on trade in asset markets. This decision is usually obtained by applying some simple rules on the stock price hierarchy.

There have been many researchers who research on technical analysis. Among them, Pring (1991), he pointed out that technical analysis is used to identify a trend reversal at the beginning, an upward trend, up to the level of confidence that the trend has reversed. Taylor and Allen (1992) show levels of technical analysis for the head of the foreign exchange sales center. Murphy (1999) discusses market action, through the use of graphs, for the purpose of forecasting future prices. Mengoli (2004) points out that a valuable momentum trading approach for the Italian stock market and he proposed the importance of behavioral theory in helping explain the profitability of technical trades. Vasiliou, Eriotis, and Papathanasiou (2006) uses the rules on MA and MACD for the Athens stock market and this rule provides strong support for choosing technical strategies. Loh (2007) compared the technical trade rules used by academics (e.g. moving average) with practitioner approach (e.g. oscillator) for five Asian countries. Lento (2006) examines the profitability of some selected technical trading rules between B & H strategies for eight stock markets in Asia-Pacific. McKenzie (2007) tested the technical trade rules for 17 selected emerging markets and concluded that no systematic trading rules could produce sufficient forecasting accuracy. Zhu and Zhou (2009) studied the primacy of the MA method from the point of view of asset allocation using S & P 500 stock data in 1926-2004. Lai, Chen, and Huang (2010) analyzed technical analysis with a psychological bias for the Taiwanese stock market and provided the disposition, cascade information, and retaining effects and each had a certain influence on the trading signal. And lastly by Metghalchi, Chang, and Garza-Gomez (2011), they analyze the rules of profitability of technical trades based on 9 popular technical indicators.

The purpose of this study was conducted with the expectation of Sutte technical analysis Indicator can be the main tool to analyze and predict future stock price and compare it with 2 other technical analysis which often used is Simple Moving Average and MACD. The object of this research is the shares that are included in LQ45 stock list that has been listed by Indonesia Stock Exchange. Analysis is done by looking at stock price movements. Stages of analysis performed are: (1) analysis of stock price movements and volume of transactions, and (2) graphical analysis and interpretation using Sutte Indicator in order to facilitate the process of reading stock price movements.

This research takes data of "A" which is listed on Indonesia Stock Exchange (IDX or BEI) in Jakarta, Indonesia. The data used in this paper is the stock price data entered into the list LQ45 period 29 November 2006 to 20 September 2016 and during that period never experienced a stock split, namely "A" in the daily period of the stock which includes the opening price (open), the highest price (high), the lowest price (low), and the closing price (close). Stock price data used also comes from the same period, i.e. from 29 November 2006 until 20 September 2016, which is obtained from the official quote of PT. Indonesia Stock Exchange (BEI) and website http://finance.yahoo.com. This stock was chosen in this study because this stock is a stock engaged in the field of crude palm oil (CPO). Trading of shares in the field of CPO is a trade that sometimes goes up sometimes down depending on the price of raw palm in the world. And stock trading in the CPO field is an interesting stock trading to discuss.

To analyze the stock movement, Sutte Indicator (SUTTE) technical analysis is used as main tool to analyze and predict stock price and as comparison, hence used technical analysis of Simple Moving Average (SMA) and Moving Average Convergence/Divergence (MACD). The formula of Sutte Indicator, SMA, and MACD as follows.
Sutte Indicator (Ahmar, 2015, 2017):

$$SUTTE\%L = \frac{C_k + C_{k-1}}{2} + C_k - L_k$$

$$SUTTE\%H = \frac{C_k + C_{k-1}}{2} + H_k - C_k$$

$$SUTTE\text{-}PRED = \frac{SUTTE\%L + SUTTE\%H}{2}$$

Where:
$C_k$ = stock's closing price on day $k$,
$C_{k-1}$ = stock's closing price on day $k-1$,
$L_k$ = the lowest price of stock's on day $k$,
$H_k$ = the highest price of stock's on the day $k$,
Sutte%L = the lowest price limit of Sutte Indicator,
Sutte%H = the highest price limit of Sutte Indicator,
Sutte-Pred = Stock prediction price using Sutte Indicator.

Simple Moving Average (SMA) (Gencay & Stengos, 1998):

$$SMA = \frac{1}{n}\sum_{i=0}^{n-1} C_{t-1}$$

Where $C_{t-1}$ = closing price of the stock at time $t-1$, and $N$ = number of days.

Moving Average Convergence/Divergence (MACD) (Panyagometh & Soonsap, 2012):

$$MACD = EMA_{short}(12-C) - EMA_{long}(26-C)$$

$$EMA_t = \alpha(C_t) + (1-\alpha)EMA_{t-1}; \alpha = \frac{2}{N+1}$$

Where $C$ = closing price of the stock, $C_t$ = closing price of the stock at time $t$, and $N$ = number of days.

To test the level of reliability of a forecasting method, so that comparison is done by comparing the level of accuracy of the forecast. The accuracy level of which is used in this study is the mean absolute deviation (MAD), mean of square error (MSE), and mean absolute percentage error (MAPE). The formula of each reliability level is as follows:

Mean Absolute Deviation (MAD)

$$\frac{\sum_{t=1}^{n}|y_t - \hat{y}_t|}{n}$$

Mean of Square Error (MSE)

$$\frac{\sum_{t=1}^{n}|y_t - \hat{y}_t|^2}{n}$$

Mean Absolute Percentage Error (MAPE)

$$\frac{\sum_{t=1}^{n}\left|\frac{(y_t - \hat{y}_t)}{y_t}\right|}{n} \times 100, (y_t \neq 0)$$

Notes: $y_t$ = Stock price day $t$, $\hat{y}_t$ = Predicted stock price day $t$-day, t = time, n = number of data.

## 3. Results and Discussion

The stock to be observed in this study are "A". This stock is the stock that most often experienced an increase and decline in stock prices. "A" stock will be conducted technical analysis process using Sutte Indicator and Simple Moving Average. The main chart for Sutte Indicator can be seen in Figure 1 below.

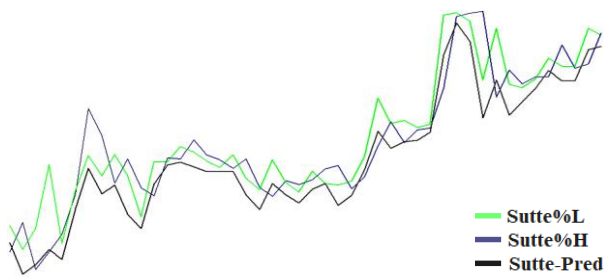

Figure 1. Main Chart of Sutte Indicator.

From Figure 1 above it is seen that SUTTE is more predictive, this is indicated from the indicator value of each analysis is related to price. It can also be compared by using reliability analysis using MSE, MAD, and MAPE which is presented in Table 1.

Table 1. Comparison of Reliability Level of SUTTE, SMA, and MACD.

| Indicator | MSE | MAD | MAPE |
|---|---|---|---|
| SUTTE | 201,329 | 121946,369 | 36.746 |
| SMA | 281,514 | 227016,351 | 51,382 |
| MACD | 784,149 | 1582877,408 | 143.122 |

In Table 1 above, it can be seen that all tested reliability levels (MAD, MSE, and MAPE), SUTTE indicators have better reliability compared to SMA and MACD. This means that in predicting the prediction accuracy level, the SUTTE Indicator can be used as a reference when compared to SMA and MACD.

In predicting the movement of stocks, SUTTE indicators have three predictive chart types, SUTTE%L, SUTTE%H, and SUTTE-PRED. These three graphs support each other to give an idea of the movement of stocks. In giving an overview of the movement of shares, between SUTTE%L and SUTTE%H mutual interconnection in the sense that if the SUTTE%L curve is above the SUTTE%H curve for a long time then this indicates that the stock price will increase and this can be a guide for the investor to buy shares and vice versa if the SUTTE%H curve is above the SUTTE%L curve then the stock price will decrease indicating that the investor should sell the shares if they do not want to get the loss. The increase and decline in stock prices is usually indicated by the intersection between the curve SUTTE%L and SUTTE%H. Illustration of the relationship between SUTTE%L and SUTTE%H can be seen in the following figure.

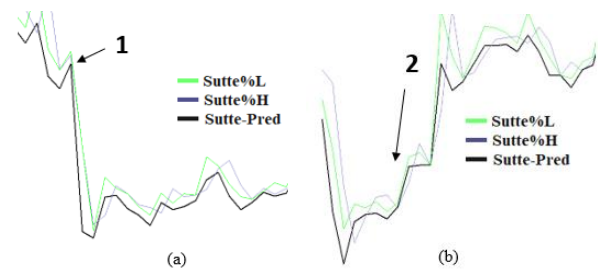

Figure 2. (a) AALI Share Movement October 27, 2015 - January 27, 2016, (b) AALI Share Movement February 24, 2016 - March 30, 2016.

In Figure 2a above, it can be seen that SUTTE%L and SUTTE%H intersect at (1) and SUTTE%H is above the SUTTE%L curve, this indicates that stock price will decrease. The intersection of this chart gives an indication to investors to take a decision that is selling its stock so that no big losses occur. Similarly in Figure 2b, it appears that SUTTE% H and SUTTE% L intersect at (2) and the SUTTE% H curve is below the SUTTE% L this informs investors that the stock price will increase and from this indication the investor can do share purchase.

## 4. Conclusions

Based on the discussion and analysis that has been done, it can be concluded that Sutte Indicator can assist investors in making decisions about stock trading and can be a reference in predicting the movement of stocks. When compared to other indicator methods (SMA and MACD) the Sutte prediction reliability level indicators have better reliability levels in terms of MSE, MAD, and MAPE.

**References**

Abbey, B. S., & Doukas, J. A. (2012). Is Technical Analysis Profitable for Individual Currency Traders. *Journal of Portfolio Management*, *1*(910), 142–150. doi:10.3905/jpm.2012.39.1.142

Ahmar, A. S. (2016, April 14). Sutte Indicator. Retrieved from https://ssrn.com/abstract=2846923

Ahmar, A. S. (2017). Sutte Indicator : A Technical Indicator in Stock Market. *International Journal of Economics and Financial Issues*, *7*(2).

Ahmar, A. S., Ahmar, A. A., Arifin, A. N. M., & Rahman, A. (2017). Predicting movement of stock of "Y" using Sutte indicator. *Cogent Economics & Finance*, 1347123. doi:10.1080/23322039.2017.1347123

Gencay, R., & Stengos, T. (1998). Moving Average Rules, Volume and the Predictability of Security Returns with Feedforward Networks. *Journal of Forecasting*, *17*, 401–414.

Han, Y., Yang, K., & Zhou, G. (2013). A New Anomaly: The Cross-Sectional Profitability of Technical Analysis. *Journal of Financial and Quantitative Analysis*, *48*(5), 1433–1461. Doi:10.1017/S0022109013000586

Lai, H. W., Chen, C. W., & Huang, C. S. (2010). Technical Analysis, Investment Psychology, and Liquidity Provision: Evidence from the Taiwan Stock Market. *Emerging Markets Finance and Trade*, *46*(5), 18–38. doi:10.2753/REE1540-496X460502

Lento, C. (2006). Tests of technical trading rules in the Asian-Pacific equity markets : A bootstrap approach. *Academy of Accounting and Financial Studies Journal*, *11*(2), 1–19. Retrieved from http://search.ebscohost.com/login.aspx?direct=true&db=bth&AN=34392104&site=ehost-live&scope=site

Li, W., & Wang, S. S. (2007). Ownership Restriction, Information Diffusion Speed, and the Performance of Technical Trading Rules in Chinese Domestic and Foreign Shares Markets. *Review of Pacific Basin Financial Markets and Policies*, *10*(4), 585–617. doi:10.1142/S0219091507001215

Loh, E. Y. L. (2007). An alternative test for weak form efficiency based on technical analysis. *Applied Financial Economics*, *17*(12), 1003–1012. doi:10.1080/09603100600749352

McKenzie, M. (2007). Technical Trading Rules in Emerging Markets and the 1997 Asian Currency Crises. *Emerging Markets Finance and Trade*, *43*(4), 46–73. doi:10.2753/REE1540-496X430403

Mengoli, S. (2004). On the source of contrarian and momentum strategies in the Italian equity market. *International Review of Financial Analysis*, *13*(3), 301–331. doi:10.1016/j.irfa.2004.02.012

Metghalchi, M., Chang, Y. H., & Garza-Gomez, X. (2011). Technical Analysis of the Taiwanese Stock Market. *International Journal of Economics and Finance*, *4*(1), 90. doi:10.5539/ijef.v4n1p90

Minitab. (2016, August 17). What are MAPE, MAD, and MSD? Retrieved from http://support.minitab.com/en-us/minitab/17/topic-library/modeling-statistics/time-series/time-series-models/what-are-mape-mad-and-msd/

Murphy, J. J. (1999). *Technical analysis of the financial markets : a comprehensive guide to trading methods and applications*. Retrieved from https://www.amazon.com

Neely, C. J., & Weller, P. A. (2011). *Technical Analysis in the Foreign Exchange Market* (No. 2011–001B). Retrieved from http://research.stlouisfed.org/wp/2011/2011-001.pdf

Nithya, J., & Thamizhchelvan, G. (2014). Effectiveness of Technical Analysis in Banking Sector of Equity Market. *IOSR Journal of Business and Management (IOSR-JBM)*, *16*(7), 20–28. Retrieved from http://www.iosrjournals.org/iosr-jbm/papers/Vol16-issue7/Version-5/C016752028.pdf

Panyagometh, K., & Soonsap, P. (2012). MACD BASED DOLLAR COST AVERAGING STRATEGY: Lessons from Long Term Equity Funds in Thailand. *Economics and Finance Review*, *2*(6), 77–84. Retrieved from http://www.businessjournalz.org/articlepdf/EFR-2606aug2(6)12g.pdf

Pring, M. J. (1991). *Technical Analysis Explained*. New York, NY: McGraw-Hill.

Suresh, A. S. (2013). A study on fundamental and technical analysis. *International Journal of Marketing, Financial Services & Management Research*, *2*(5), 44–59.

Taylor, M. P., & Allen, H. (1992). The use of technical analysis in the foreign exchange market. *Journal of International Money and Finance*, *11*(3), 304–314. doi:10.1016/0261-5606(92)90048-3

Vasiliou, D., Eriotis, N., & Papathanasiou, S. (2006). How rewarding is technical analysis? Evidence from Athens Stock Exchange. *Operational Research*, *6*(2), 85–102. doi:10.1007/BF02941226

Zhu, Y., & Zhou, G. (2009). Technical analysis: An asset allocation perspective on the use of moving averages. *Journal of Financial Economics*, *92*(3), 519–544. doi:10.1016/j.jfineco.2008.07.002